\documentclass[prb,twocolumn,preprintnumbers,amsmath,amssymb]{revtex4-1}
\usepackage{amsmath}
\usepackage{amssymb}
\usepackage{mathrsfs}
\usepackage{graphicx}
\usepackage{dcolumn}
\usepackage{bm}
\usepackage{color}
\usepackage{hhline}
\usepackage{rotating}

\begin{document}

\title{High-resolution combined tunneling electron charge and spin transport theory of N\'eel and Bloch skyrmions}

\author{Kriszti\'an Palot\'as$^{1,2,3,4,}$}
\email{palotas@phy.bme.hu}

\affiliation{1 Slovak Academy of Sciences, Institute of Physics, Department of Complex Physical Systems, Center for Computational Materials Science, D\'ubravsk\'a cesta 9, SK-84511 Bratislava, Slovakia\\
2 University of Szeged, MTA-SZTE Reaction Kinetics and Surface Chemistry Research Group, Rerrich B.\ t\'er 1, H-6720 Szeged, Hungary\\
3 Budapest University of Technology and Economics, Department of Theoretical Physics, Budafoki \'ut 8, H-1111 Budapest, Hungary\\
4 Hungarian Academy of Sciences, Wigner Research Center for Physics, Institute for Solid State Physics and Optics, P.\ O.\ Box 49, H-1525 Budapest, Hungary}

\date{\today}

\begin{abstract}

Based on a combined charge and vector spin transport theory capable of imaging noncollinear magnetic textures on surfaces with spin-polarized scanning tunneling microscopy (SP-STM), the high-resolution tunneling electron charge and coupled spin transport properties of a variety of N\'eel- and Bloch-type skyrmions are investigated. Axially symmetric skyrmions are considered within the same topology class characterized by a vorticity value of $m=1$, and their helicities are varied by taking $\gamma=0$ and $\pi$ values for the N\'eel skyrmions and $\gamma=-\pi/2$ and $\pi/2$ values for the Bloch skyrmions. Depending on the orientation of the magnetization of the STM tip as well as on the helicity and the time-reversal of the skyrmionic spin structures, several relationships between their spin transport vector components, the in-plane and out-of-plane spin transfer torque and the longitudinal spin current, are identified. The magnitudes of the spin transport vector quantities show close relation to standard charge current SP-STM images. It is also demonstrated that the SP-STM images can be used to determine the helicity of the skyrmions. Moreover, the modified spin polarization vectors of the conduction electrons due to the local chirality of the complex spin texture are incorporated into the tunneling model. It is found that this effect modifies the apparent size of the skyrmions. These results contribute to the proper identification of topological surface magnetic objects imaged by SP-STM, and deliver important parameters for current-induced spin dynamics.

\end{abstract}

\maketitle

\section{Introduction}
\label{sec_int}

Magnetic skyrmions are not only artistic real-space spin textures but they show considerable potential for being important building blocks for future technologies. The topological protection of their spin structures arises from the looping of magnetic domain walls of various complexity \cite{Okubo,Leonov,Lin,zhang16prb,rozsa-sk3,yang17mobius,hagemeister18kpi_sk}, and they are usually characterized by topological invariants, like the the topological charge also known as winding number \cite{Leonov,nagaosa13,leonov16}, or the vorticity \cite{Lin,rozsa-sk3,nagaosa13}. Another quantitative characteristic of skyrmions, the helicity \cite{nagaosa13,yu12} does not change their topological classification \cite{rozsa-sk3}, but its dynamics show complex features \cite{diaz16,zhang17,ritzmann18}. On the other hand, the reversal of the external magnetic field resulting in the time-reversal of all the spins rotates the helicity by $\pi$ and reverts the sign of the topological charge, while keeping the vorticity unchanged \cite{rozsa-sk3}. Localized skyrmions are very promising for future technological use in magnetic data storage, information carrier and spintronic devices \cite{nagaosa13,fert13,zhang15,wiesendanger16natrevmat,fert17natrevmat} due to their energetically much favored transport compared to domain walls \cite{iwasaki13,jonietz10}.

The formation of magnetic skyrmions is governed by a subtle interplay of various magnetic exchange interaction and anisotropy terms. First-principles calculations contribute crucially to the understanding of the formation of skyrmionic structures in thin film systems \cite{heinze11,dupe14,simon14,rozsa-sk2}. For example, in most of the cases the preferred chirality of the skyrmions in thin films is due to the emergence of the antisymmetric Dzyaloshinsky-Moriya interaction (DMI) \cite{Dzyaloshinsky,Moriya} at the magnetic layer with broken inversion symmetry (interfacial DMI). The competition between the strength of the DMI vectors, which tend to rotate the spin moments in the magnetic layer, and the isotropic scalar Heisenberg exchange interactions and the magnetic anisotropy, preferring collinear spin moments, determines the complexity of the formed real-space magnetic textures. Another example, which contributes to this picture, is the frustrated Heisenberg exchange interactions that turned out to be important for the stabilization of localized skyrmionic spin configurations with different topologies \cite{Okubo,Leonov,Lin,rozsa-sk3}. Moreover, relevant for practical applications, the temperature effects on the skyrmion stability have been experimentally \cite{moreau16,boulle16,woo16} and theoretically \cite{hagemeister15,rozsa-sk1,bessarab15,lobanov16,stosic17,bessarab18} investigated.

Spin-polarized scanning tunneling microscopy (SP-STM) is a convenient tool to study magnetic skyrmions and other magnetic objects at surfaces \cite{bergmann14,romming15prl,palotas17prb,wiesendanger09review}. The real-space spin textures can be imaged in high spatial resolution by probing them with low energy electrons tunneling between the noncollinear magnetic sample and the STM tip. Here, the key challenge is the correct identification of the topological objects based on measured SP-STM image contrasts \cite{palotas17prb}. At certain conditions involving higher energy tunneling electrons the skyrmions are modified due to localized spin-polarized currents introduced by the STM tip, and controlled creation and annihilation of the skyrmions have been demonstrated \cite{romming13,hsu17elec} using local current pulses with opposite voltage polarities. Although the annihilation of skyrmions has recently been extensively studied by using minimum energy path calculations \cite{bessarab15,lobanov16,stosic17,bessarab18}, the detailed microscopic insight into the role of the localized spin transfer due to the tunneling electrons in these processes is yet to be uncovered. Thus, another key challenge is how to gain local information on the tunneling spin transport properties of complex surface magnetic objects in SP-STM junctions.

The present work contributes to the solution of both key challenges. Taking N\'eel- and Bloch-type skyrmions within the same topology class, with cycloidal and helical domain walls, respectively, a comparison of the tunneling charge and spin transport characteristics of eight distinct skyrmions is provided. An electron tunneling theory for the consistently combined description of charge and vector spin transport in magnetic STM junctions within the three-dimensional (3D) Wentzel-Kramers-Brillouin (WKB) framework \cite{palotas16prb,palotas18stt1} is employed. It is shown that the charge current SP-STM images can be used to determine the helicity of the skyrmions. Moreover, current-induced tunneling spin transport quantities, the longitudinal spin current (LSC) vector, the spin transfer torque (STT) vector, and their magnitudes are calculated in high spatial resolution. Interesting relationships among the LSC vectors as well as among the STT vectors and their in-plane and out-of-plane components are identified depending on the helicities of the skyrmions and on the time-reversal symmetry of the spin structures. The connections between the charge current SP-STM image contrasts and the magnitudes of the LSC and the STT are also pointed out. Therefore, the SP-STM contrast behaviors of the charge current, e.g., reported in Ref.\ \onlinecite{palotas17prb}, can be directly transferred to the spin transport magnitudes.

The paper is organized as follows. In section \ref{sec_meth} the combined tunneling electron charge and vector spin transport 3D-WKB theoretical model in magnetic STM is presented, and tunneling parameters are described. The inclusion of the effect of the noncollinear spin texture on the spin polarization of conduction electrons, and discussions on the context and limitations of the model and on the possibility of including other (spin-orbit-)torque terms are also provided. The studied eight skyrmions are introduced in section \ref{sec_sky}. The effect of the complex spin texture on the spin polarization of conduction electrons is analyzed in section \ref{sec_ncmr}. The electron charge (charge current) and vector spin transport (longitudinal spin current and spin transfer torque) properties of the skyrmions are investigated in section \ref{sec_chtr} and \ref{sec_spintr}, respectively. Summary and conclusion are found in section \ref{sec_conc}.

\section{Method and systems}

\subsection{3D-WKB electron tunneling theory}
\label{sec_meth}

For the description of the tunneling electron charge and spin transport properties of the skyrmions, the 3D-WKB theory is used. The method is based on the original idea of Tersoff and Hamann \cite{tersoff83,tersoff85}, where the tunneling current is obtained as the superposition of one-dimensional (1D) WKB electron charge transport contributions between the surface atoms and the apex of the STM tip. This method was generalized for noncollinear magnetic surfaces by Heinze \cite{heinze06}, and the SP-STM imaging of complex magnetic surface textures became possible, employing the tunneling magnetoresistance (TMR) effect. This approach has been used to extract the real-space magnetic structure of distorted skyrmions from a series of experimentally obtained SP-STM images \cite{hsu17elec}, and theoretical calculations identified the magnetic objects \cite{hagemeister16}. Further developments of the tunneling theory included the explicit consideration of the electronic structure of the tip in 3D-WKB (spin-polarized) STM \cite{palotas11stm,palotas12orb,palotas13contrast,mandi14fe,nita14} and tunneling spectroscopy \cite{palotas11sts,palotas12sts}, and an enhanced parameter space for studying tip geometry effects on the STM contrast \cite{mandi13tiprot,mandi14rothopg,mandi15tipstat}. An atom-superposition approach to include the spin-orbit-coupling(SOC)-related tunneling anisotropic magnetoresistance (TAMR) effect on the atomic scale was presented in Ref.\ \onlinecite{bergmann12}.

The developed methods so far excluded the effect of the noncollinear spin texture on the spin polarization of conduction electrons. It was shown that the non-coplanarity (chirality) of the spins results in the emergence of persistent in-plane electric currents \cite{tatara04}. These, in effect, produce a magnetic field \cite{nagaosa13,dias16} that alters the direction of the spin polarization of conduction electrons from the local exchange field. In the present work the inclusion of this effect into the 3D-WKB SP-STM theory is introduced.

Beside the spin-polarized charge current, the calculation of tunneling spin transport quantities, the LSC and the STT, was first proposed taking collinear magnets in the magnetic tunnel junction \cite{palotas16prb}. The extension to noncollinear magnetic surfaces delivered important insights to the high-resolution charge transport properties of skyrmionic objects with various topologies \cite{palotas17prb} and to the high-resolution vector spin transport characteristics of an individual skyrmion \cite{palotas18stt1}. The 3D-WKB electron tunneling theory is implemented in the 3D-WKB-STM code \cite{palotas14fop}.

The charge current ($I$) \cite{heinze06} and the current-induced ingredients for the LSC vectors ($\mathbf{T}^{L}$) and the STT vector components (in-plane $\mathbf{T}^{\parallel}$ and out-of-plane $\mathbf{T}^{\perp}$) at the magnetic tip apex position $\mathbf{R}_{T}$ (characterized by the spin unit vector $\mathbf{s}_T$) are given by the superposition of atomic contributions (sum over "$a$") from the sample surface localized spin unit vectors $\mathbf{s}_S^a$ at positions $\mathbf{R}_a$ in the limits of elastic tunneling and low bias voltage $V$ as \cite{palotas18stt1}
\begin{eqnarray}
I(\mathbf{R}_{T})&=&\frac{e^2}{2\pi\hbar}|V|\sum_a h(\mathbf{R}_{T}-\mathbf{R}_a)(1+P_{S}P_{T}\cos\phi_a),\nonumber\\
\mathbf{T}^{TL}(\mathbf{R}_{T})&=&e|V|\sum_a h(\mathbf{R}_{T}-\mathbf{R}_a)(P_{T}+P_{S}\cos\phi_a)\mathbf{s}_T,\nonumber\\
\mathbf{T}^{SL}(\mathbf{R}_{T})&=&e|V|\sum_a h(\mathbf{R}_{T}-\mathbf{R}_a)(P_{S}+P_{T}\cos\phi_a)\mathbf{s}_S^a,\nonumber\\
\mathbf{T}^{T\parallel}(\mathbf{R}_{T})&=&e|V|\sum_a h(\mathbf{R}_{T}-\mathbf{R}_a)P_{S}\mathbf{s}_T\times(\mathbf{s}_S^a\times\mathbf{s}_T),\nonumber\\
\mathbf{T}^{S\parallel}(\mathbf{R}_{T})&=&e|V|\sum_a h(\mathbf{R}_{T}-\mathbf{R}_a)P_{T}\mathbf{s}_S^a\times(\mathbf{s}_T\times\mathbf{s}_S^a),\nonumber\\
\mathbf{T}^{\perp}(\mathbf{R}_{T})&=&e|V|\sum_a h(\mathbf{R}_{T}-\mathbf{R}_a)P_{S}P_{T}\mathbf{s}_S^a\times\mathbf{s}_T.\label{Eq_transport}
\end{eqnarray}
Here, $e$ is the elementary charge and $\hbar$ is the reduced Planck constant. The upper indices $T$ or $S$ denote the tip or sample side, on which spin moments the LSC and the in-plane STT are acting. $\phi_a$ is the angle between the spin moments of surface atom "$a$" and the tip apex atom, thus $\cos\phi_a=\mathbf{s}_S^a\cdot\mathbf{s}_T$. $P_S$ and $P_T$ denote the spin polarization of the surface atoms and the tip apex atom at their respective Fermi energies, and site-dependent $P_S^a$ can be considered as well. Note that the effective spin polarization ($P_{\mathrm{eff}}=P_SP_T$) only enters the charge current and the out-of-plane torque expressions, and not the longitudinal spin current and the in-plane torque. This means that $P_{\mathrm{eff}}$ is not sufficient to characterize the spin polarization of the magnetic tunnel junction concerning spin transport quantities, and $P_S$ and $P_T$ are independent parameters in the tunneling model \cite{palotas18stt1}.

The electron transmission function is \cite{heinze06}
\begin{equation}
h(\mathbf{r})=\exp\left[-\sqrt{8m\Phi/\hbar^2}|\mathbf{r}|\right]
\label{Eq_transmission}
\end{equation}
with $m$ the electron's mass and $\Phi$ the effective work function. In the transmission all electron states are considered as exponentially decaying spherical states \cite{tersoff83,tersoff85,heinze06}, and their orbital-dependence is omitted for simplicity. It is, in principle, possible to take into account the orbital-dependence of the electronic states in the tunneling transmission based on {\it{ab initio}} calculations \cite{palotas12orb,mandi13tiprot,mandi14fe,mandi14rothopg,mandi15tipstat,palotas16prb}, and this is expected to provide a better description of the electron charge and spin tunneling process at larger bias voltages, which is, however, not in the focus of the present study. Due to the fast decay of the transmission function, in the following discussion the spin direction $\mathbf{s}_S^a$ of the surface atom closest to the tip apex position $\mathbf{R}_{T}$ is understood when referring to a single $\phi_a$ value \cite{palotas18stt1}.

While the absolute charge current $I(\mathbf{R}_{T})$ in Eqs.~(\ref{Eq_transport}) is independent of the direction of the electron tunneling, the spin transport vectors strongly depend on that. In agreement with conventional spin transport interpretations \cite{ralph08}, the LSC and the in-plane STT vectors change sign and the out-of-plane STT vector does not change sign by reversing the bias polarity \cite{palotas18stt1}, which is by convention $V>0$ at tip $\rightarrow$ sample ($T\rightarrow S$), and $V<0$ at sample $\rightarrow$ tip ($S\rightarrow T$) tunneling. The dependences of the spin transport vectors on the direction of the electron tunneling in relation to Eqs.~(\ref{Eq_transport}) are summarized below:
\begin{eqnarray}
\mathbf{T}^{T\rightarrow S,TL}(\mathbf{R}_{T})&=&-\mathbf{T}^{S\rightarrow T,TL}(\mathbf{R}_{T})=\mathbf{T}^{TL}(\mathbf{R}_{T}),\nonumber\\
\mathbf{T}^{T\rightarrow S,SL}(\mathbf{R}_{T})&=&-\mathbf{T}^{S\rightarrow T,SL}(\mathbf{R}_{T})=\mathbf{T}^{SL}(\mathbf{R}_{T}),\nonumber\\
\mathbf{T}^{T\rightarrow S,T\parallel}(\mathbf{R}_{T})&=&-\mathbf{T}^{S\rightarrow T,T\parallel}(\mathbf{R}_{T})=\mathbf{T}^{T\parallel}(\mathbf{R}_{T}),\nonumber\\
\mathbf{T}^{T\rightarrow S,S\parallel}(\mathbf{R}_{T})&=&-\mathbf{T}^{S\rightarrow T,S\parallel}(\mathbf{R}_{T})=\mathbf{T}^{S\parallel}(\mathbf{R}_{T}),\nonumber\\
\mathbf{T}^{T\rightarrow S,\perp}(\mathbf{R}_{T})&=&\mathbf{T}^{S\rightarrow T,\perp}(\mathbf{R}_{T})=\mathbf{T}^{\perp}(\mathbf{R}_{T}),\nonumber\\
\mathbf{T}^{T\rightarrow S,T}(\mathbf{R}_{T})&=&\mathbf{T}^{\perp}(\mathbf{R}_{T})+\mathbf{T}^{T\parallel}(\mathbf{R}_{T}),\nonumber\\
\mathbf{T}^{S\rightarrow T,T}(\mathbf{R}_{T})&=&\mathbf{T}^{\perp}(\mathbf{R}_{T})-\mathbf{T}^{T\parallel}(\mathbf{R}_{T}),\nonumber\\
\mathbf{T}^{T\rightarrow S,S}(\mathbf{R}_{T})&=&\mathbf{T}^{\perp}(\mathbf{R}_{T})+\mathbf{T}^{S\parallel}(\mathbf{R}_{T}),\nonumber\\
\mathbf{T}^{S\rightarrow T,S}(\mathbf{R}_{T})&=&\mathbf{T}^{\perp}(\mathbf{R}_{T})-\mathbf{T}^{S\parallel}(\mathbf{R}_{T}).\label{Eq_transport2}
\end{eqnarray}
Here, the total STT vectors $\mathbf{T}^{T\rightarrow S,T}$ and $\mathbf{T}^{S\rightarrow T,T}$ act on the spin moment of the tip apex atom at $T\rightarrow S$ and $S\rightarrow T$ tunneling, respectively. Similarly, the total STT vectors $\mathbf{T}^{T\rightarrow S,S}$ and $\mathbf{T}^{S\rightarrow T,S}$ act on the spin moments of the sample surface at the indicated tunneling directions.

The TAMR \cite{bergmann12}, the noncollinear magnetoresistance (NCMR) \cite{hanneken15,kubetzka17}, and their combined tunneling spin mixing magnetoresistance (TXMR) \cite{crum15} effects were previously studied by employing non-magnetic tips only. While the spin noncollinearity can have a first order effect on the electronic structure, the SOC typically has a second order effect \cite{crum15}. Employing magnetic STM tips it has to be noted that the charge current contrast formation and the coupled spin transport properties are dominated by the TMR effect. Nonetheless, in the following the inclusion of the noncollinearity of the spins into the 3D-WKB method employing magnetic tips is proposed. This way, insights into the origin of the NCMR can be obtained, that is a non-local effect arising from the magnetic environment of the complex spin texture, and consequently the spin polarization of conduction electrons is modified. The scalar non-coplanarity (chirality) of the spin at site "$a$" can be calculated as
\begin{equation}
C_a=\frac{1}{4\pi}\frac{1}{6}\mathbf{s}_S^a\cdot\sum_{i,j}\mathbf{s}_S^i\times\mathbf{s}_S^j,
\label{Eq_chirality}
\end{equation}
where "$i$" and "$j$" are selected from the 6 neighboring spins of site $"a"$ forming equilateral triangular plaquettes with "$a-i-j$" counter-clockwise order, and the averaging provides numerically better results, see also Ref.\ \onlinecite{hu17} for a similar summation over 3 triangular plaquettes. The spin chirality $C_a$ equals the topological charge density in the continuum limit \cite{berg81}, thus $\sum_aC_a=Q$ is the topological charge, see section \ref{sec_sky}. The emergent (dimensionless) magnetic field at site "$a$" is
\begin{equation}
\mathbf{B}_a=B_0 C_a \mathbf{e}_z,
\label{Eq_field}
\end{equation}
and thus the direction of the spin polarization of conduction electrons at site "$a$" can be obtained as $\mathbf{s'}_S^a=||\mathbf{s}_S^a+\mathbf{B}_a||$, where $||.||$ means normalization to unit vector. This $\{\mathbf{s'}_S^a\}$ set of modified spin polarization directions should be used in Eqs.~(\ref{Eq_transport}) instead of the set of $\{\mathbf{s}_S^a\}$ local spin moment directions when the chirality of the spin texture is accounted for. This effect is studied on a selected skyrmion in section \ref{sec_ncmr}.

Even though the SOC typically has a second order effect on the electronic structure \cite{crum15}, the inclusion of spin-orbit torques (SOTs) into the 3D-WKB electron tunneling model is, in principle, also possible. This can be achieved in two ways:\\
(i) by calculating the atomic site-dependent modified spin polarization of conduction electrons due to effective magnetic fields corresponding to the system-dependent SOC Hamiltonian \cite{gambardella11}, similarly as described in the previous paragraph for the chirality of the spin texture;\\
(ii) if the SOT coefficients are known from another source (or taken as parameters) then the scalar coefficients of the torque vector formulas in Eqs.~(\ref{Eq_transport}) have to be transformed as STT$\rightarrow$(STT+SOT), e.g., $\mathbf{T}^{T\parallel}:\;P_S\rightarrow P'_S$, $\mathbf{T}^{S\parallel}:\;P_T\rightarrow P'_T$, and $\mathbf{T}^{\perp}:\;(P_SP_T)\rightarrow (P_SP_T)'$. Here, the coupled longitudinal spin-orbit currents (LSOCs) have to be properly defined, and the SOC-related TAMR\cite{bergmann12} in the charge current has to be revised.\\
Note, however, that the inclusion of SOT through effective SOC fields, and a consistent theory of the charge current, STT, SOT, and related LSC and LSOC are beyond the scope of the present paper, and neither orbital angular momentum transfer nor spin-orbit coupling in the spin transfer are considered, only current-induced spin transfer within a combined charge current-LSC-STT SP-STM theory \cite{palotas18stt1}. This limitation has to be taken into account when interpreting the presented torque results that are clearly spin transfer torques due to tunneling electrons.

During the simulations employing the above electron tunneling model, the following computational parameters were used: the absolute bias voltage is $|V|=1.5$ meV, and the effective work function is $\Phi=5$ eV. Motivated by the reported electronic structure of a recent work \cite{crum15}, $P_S=-0.5$ is chosen. $P_T$ is set to $-0.8$ to arrive at an effective spin polarization $P_{\mathrm{eff}}=+0.4$, which value was also taken in related works on SP-STM of skyrmionic spin structures with different topologies \cite{palotas17prb,dupe16stm} and on studying spin transport properties of a single skyrmion \cite{palotas18stt1}. For investigating the effect of the chirality of the noncollinear spin texture on the spin polarization of conduction electrons, the parameters $B_0=\pm 75$ in Eq.~(\ref{Eq_field}) were selected.

SP-STM images of the charge current are shown in constant current mode using a white-brown-black color palette corresponding to maximum-medium-minimum apparent heights. Employing the reported parameters, the current value $I=10^{-4}$ nA of the constant-current contours corresponds to about 6 \AA\;minimal tip-sample distance and corrugation values between 30 and 40 pm \cite{palotas17prb}. The spin transport quantities (LSC and STT vectors and scalar magnitudes) are given in constant-height mode at 6 \AA\;tip-sample distance. The magnitudes of the LSC and the STT are shown using a red-green-blue color palette corresponding to maximum-medium-minimum values of the individual images. Although the LSC and the STT vectors are calculated in the same high resolution (1 \AA) as the charge current and the magnitudes of the LSC and the STT, the vector spin transport quantities are reported with a lateral resolution of 5 \AA\;for visualization reasons.

\subsection{Skyrmionic spin structures}
\label{sec_sky}

To utilize the above-described combined tunneling charge and vector spin transport 3D-WKB theory of noncollinear magnetic surfaces, a set of skyrmions is considered and their charge and spin transport properties in high spatial resolution are investigated.

\begin{figure*}[t]
\begin{tabular}{c}
\includegraphics[width=1.0\textwidth,angle=0]{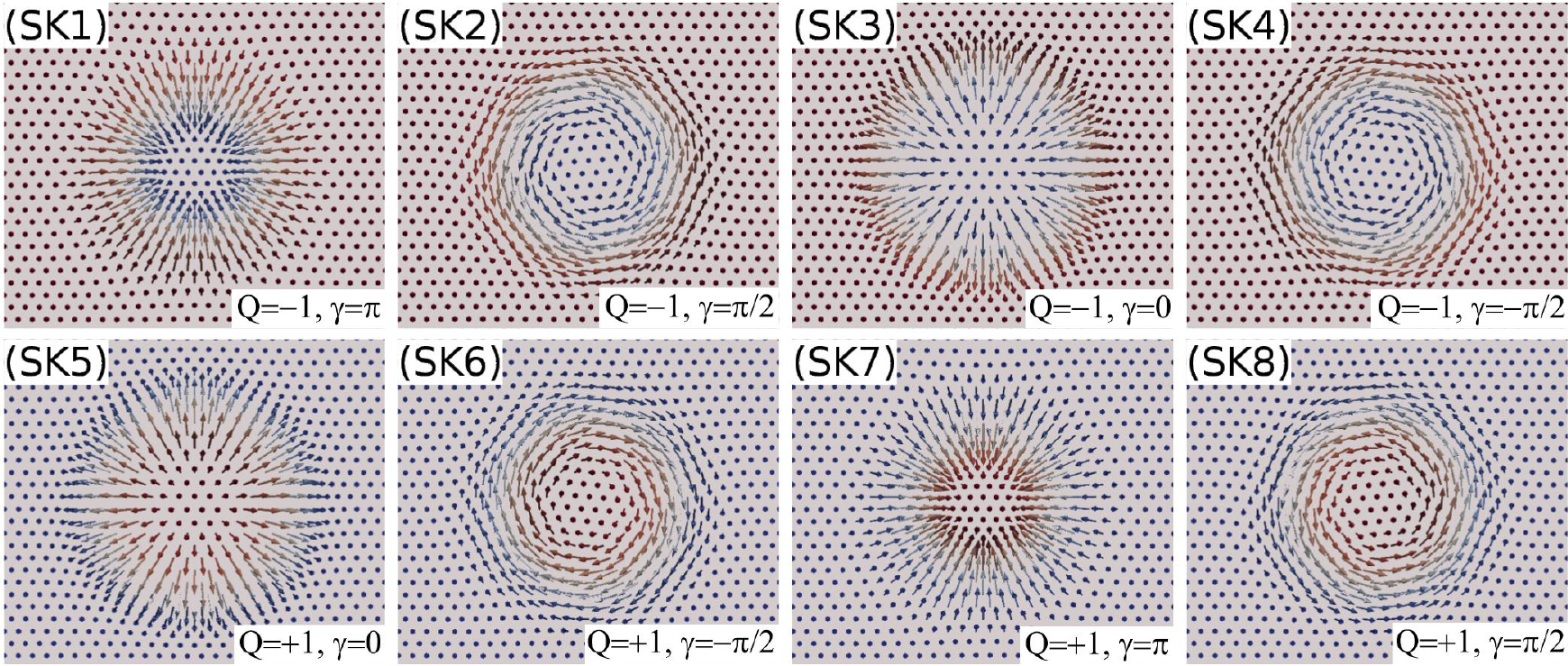}
\end{tabular}
\caption{\label{Fig1} Spin structures of skyrmions with vorticity $m=1$ and various topological charges ($Q$) and helicities ($\gamma$) explicitly shown. (SK1,SK3,SK5,SK7) N\'eel-type, (SK2,SK4,SK6,SK8) Bloch-type skyrmions. Red and blue colors respectively correspond to positive and negative out-of-plane ($z$) spin components.}
\end{figure*}

The localized skyrmionic (classical) spin configurations of the sample surface can be written in a continuum description as $\mathbf{s}_S^a=\mathbf{s}(r,\varphi)$, where $\mathbf{s}$ is a unit vector and $(r,\varphi)$ are polar coordinates of the two-dimensional surface plane. The localized spins can be represented as $\mathbf{s}(r,\varphi)=[\sin\Theta(r)\cos\Phi(\varphi),\sin\Theta(r)\sin\Phi(\varphi),\cos\Theta(r)]$ in the circular approximation \cite{Lin,nagaosa13}. In this case
\begin{equation}
\Phi(\varphi)=m\varphi+\gamma,
\end{equation}
where $m$ is the vorticity and $\gamma$ is the helicity. The vorticity $m$ is a topological quantity: it expresses how many times and in which direction the in-plane component of the spins rotates around the circle when following a closed curve involving the center of the spin structure. A more frequently used quantity to describe topological states, the topological charge $Q$ or winding number counts how many times the vector field $\mathbf{s}$ winds around the unit sphere: $Q=\frac{1}{4\pi}\int\mathbf{s}\cdot(\partial_x\mathbf{s}\times\partial_y\mathbf{s})dxdy$, where the surface integral has to be performed over the area of the localized spin structure. The topological charge $Q$ is related to the vorticity $m$ as \cite{rozsa-sk3,palotas17prb}: $Q=-[\cos\Theta(r)]^{\infty}_{0}m/2=-\mathrm{sgn}(B)m$, thus $Q$ is equal to $-m$ or $m$ depending on the direction of the external magnetic field $\mathbf{B}$. Here, $\mathrm{sgn}(B)=1$ corresponds to $\Theta(r\rightarrow\infty)=0$, i.e., the spins far from the localized skyrmionic structure point outwards from the surface (in the $+z$ direction) that is the direction of the external magnetic field $\mathbf{B}$, and $\Theta(r=0)=\pi$ considering single-domain skyrmionic structures. On the other hand, $\mathrm{sgn}(B)=-1$ means $\Theta(r\rightarrow\infty)=\pi$, i.e., the spins far from the localized skyrmionic structure point inwards to the surface (in the $-z$ direction) parallel to the external magnetic field $\mathbf{B}$, and $\Theta(r=0)=0$ for single-domain skyrmionic structures.

The helicity $\gamma$ can, in principle, be continuously changed that prescribes $\Phi(\varphi=0)=\gamma$. In realistic surface magnetic systems the presence of the interface DMI due to the break of inversion symmetry restricts the choice of $\gamma$, and it was shown in Ref.\ \onlinecite{rozsa-sk3} that only the $m=1$ vorticity value provides spin structures having axial symmetry, exactly fulfilling the circular approximation. For $m=1$ the helicity is well-defined, and the spin structures with $\gamma=0$ or $\pi$ are called N\'eel-type skyrmions and with $\gamma=-\pi/2$ or $\pi/2$ are called Bloch-type skyrmions. The continuous tuning of the helicity has been proposed between two limiting cases, the N\'eel-skyrmions (Rashba limit) and the Bloch-skyrmions (Dresselhaus limit), by tuning the ratio of the interface and bulk DMI strengths \cite{rowland16}. N\'eel-skyrmions are hedgehog-like spin structures with cycloidal domain wall and they are preferred at strong interface DMI, while Bloch-skyrmions are vortex-like spin structures with helical domain wall and they are preferred at strong bulk DMI.

The focus of the present work is to investigate the tunneling spin transport properties and their relation to the tunneling charge transport of various skyrmions. For simplicity, the spin structures are restricted to belong to the same topology class fulfilling the circular approximation and having axial symmetry. Therefore, a set of skyrmions with vorticity $m=1$ are taken, which means that the in-plane component of the spins is rotating in the same direction making exactly $2\pi$ rotation when moving along a closed curve involving the center of the localized spin structure. Following the above considerations, eight spin structures can be defined as shown in Figure \ref{Fig1}. These correspond to two opposite directions of the external magnetic field: $\mathrm{sgn}(B)=1$ and $Q=-1$ (SK1-4) and $\mathrm{sgn}(B)=-1$ and $Q=1$ (SK5-8), combined with four different helicities: $\gamma=\pi$ (SK1, SK7), $\gamma=\pi/2$ (SK2, SK8), $\gamma=0$ (SK3, SK5), and $\gamma=-\pi/2$ (SK4, SK6). The initial spin structure was SK1 with $Q=-1$ and $\gamma=\pi$, which was obtained by a combination of {\it{ab initio}} and spin dynamics calculations on a (111)-oriented surface with $C_{3v}$ crystallographic symmetry \cite{rozsa-sk2,rozsa-sk3}. The rest of the skyrmions with $Q=-1$ (first row of Fig.\ \ref{Fig1}) were calculated by rotating the in-plane spin components of SK1 by $-\pi/2$, $\pi$, and $\pi/2$ for SK2, SK3, and SK4, respectively. The spin configurations with $Q=1$ (second row of Fig.\ \ref{Fig1}) were obtained by correspondingly time-reversing the spins in the first row of Fig.\ \ref{Fig1}. This transformation results in the reversal of the external magnetic field (and thus the sign of the topological charge) and also to a rotation of the helicity by $\pi$, see the time-reversed spin structure pairs in Fig.\ \ref{Fig1}: SK1-SK5, SK2-SK6, SK3-SK7, and SK4-SK8.

\section{Results and discussion}
\label{sec_res}

\subsection{Effect of the complex spin texture on the spin polarization of conduction electrons}
\label{sec_ncmr}

\begin{figure*}[t]
\begin{tabular}{c}
\includegraphics[width=1.0\textwidth,angle=0]{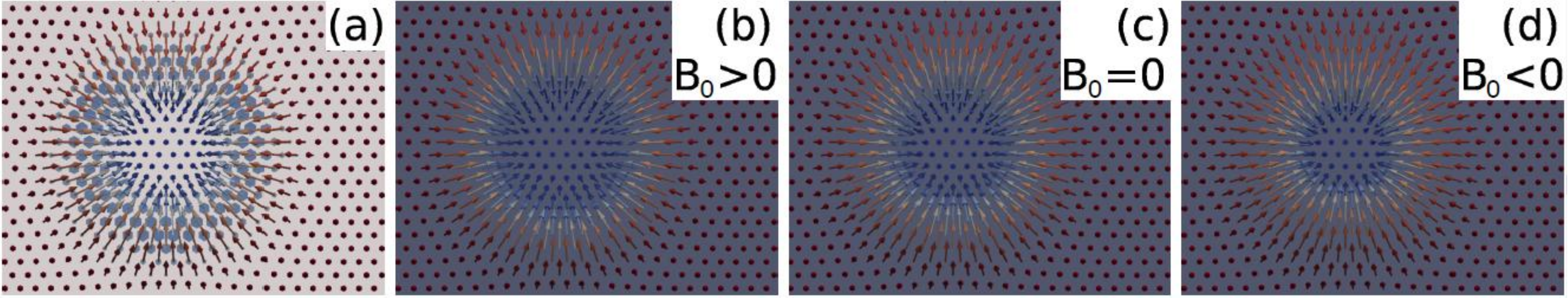}
\end{tabular}
\caption{\label{Fig2}(a) Spin structure $\{\mathbf{s}_S^a\}$ of the SK1 skyrmion and the emergent magnetic field $\mathbf{B}_a$ at atomic positions (enhanced light blue arrows perpendicular to the surface plane, $B_0>0$ in Eq.~(\ref{Eq_field})) due to the calculated chirality $C_a$ (Eq.~(\ref{Eq_chirality})) of the spin texture $\{\mathbf{s}_S^a\}$. (b)-(d) Modified spin polarization directions $\{\mathbf{s'}_S^a\}$ of conduction electrons due to the spin chirality with $B_0=75$ (b), $B_0=0$ (c), and $B_0=-75$ (d) coefficients in Eq.~(\ref{Eq_field}). Red and blue colors respectively correspond to positive and negative out-of-plane ($z$) vector components.}
\end{figure*}

First, the effect of the noncollinear spin texture on the spin polarization of conduction electrons is investigated, taking the SK1 skyrmion from Fig.\ \ref{Fig1}. The original spin structure $\{\mathbf{s}_S^a\}$ of the SK1 skyrmion together with the emergent magnetic field $\mathbf{B}_a$ (Eq.~(\ref{Eq_field})) due to the local spin chirality $C_a$ (Eq.~(\ref{Eq_chirality}), a discretized topological charge density) at atomic positions are shown in Fig.\ \ref{Fig2}(a). The distribution of $\mathbf{B}_a$, that point toward the surface (negative $z$ components, light blue color), indeed correlates very well with those lattice sites, where the topological charge density is non-zero, see, e.g., Fig.\ 4(a) of Ref.\ \onlinecite{palotas17prb}, and such sites form a ring around the skyrmion center. The modified spin polarization directions of conduction electrons $\{\mathbf{s'}_S^a\}$ due to $\mathbf{B}_a$ are shown in Figs.\ \ref{Fig2}(b)-(d), taking three different $B_0$ coefficients in Eq.~(\ref{Eq_field}). An apparent shrinking of the skyrmion size with the change of the $B_0$ coefficients from positive to negative values is evident following the sequence of Figs.\ \ref{Fig2}(b)-(d). Thus, the chirality of the spin texture results in apparent size differences of $\{\mathbf{s'}_S^a\}$ (Figs.\ \ref{Fig2}(b),(d)) compared to the original SK1 skyrmion $\{\mathbf{s}_S^a\}$ in Fig.\ \ref{Fig2}(a), or $\{\mathbf{s'}_S^a\}=\{\mathbf{s}_S^a\}$ ($B_0=0$) in Fig.\ \ref{Fig2}(c). Note that similar apparent size differences are obtained for the other skyrmions SK2-SK8 from Fig.\ \ref{Fig1} (not shown). This is due to the fact that all SK1-SK8 skyrmions belong to the same topology class characterized by a vorticity value of $m=1$, and their topological charge density and local spin chirality distributions on the lattice sites are respectively the same for the time-reversed SK1-4 and SK5-8 sets. Such a demonstrated apparent skyrmion size effect through the modified spin polarization directions of conduction electrons due to the noncollinearity of the spins should be considered when determining actual skyrmion sizes from SP-STM measurements and related material-specific magnetic interaction and anisotropy (micromagnetic) parameters \cite{romming15prl}. The magnitude and direction of this size effect clearly depend on the absolute value and sign of the $B_0$ coefficients in Eq.~(\ref{Eq_field}).

Note that {\emph{except the skyrmion size}} the calculated tunneling charge and coupled spin transport properties of the skyrmions in sections \ref{sec_chtr} and \ref{sec_spintr} are expected to be qualitatively unaffected since their topological properties do not change, and in the remaining of the paper the spin polarization directions of the conduction electrons correspond to the actual spin structures taking $B_0=0$, just as shown in Fig.\ \ref{Fig2}(c) for the SK1 skyrmion.

\subsection{Charge transport characteristics of the skyrmions}
\label{sec_chtr}

\begin{figure*}[t]
\begin{tabular}{c}
\includegraphics[width=1.0\textwidth,angle=0]{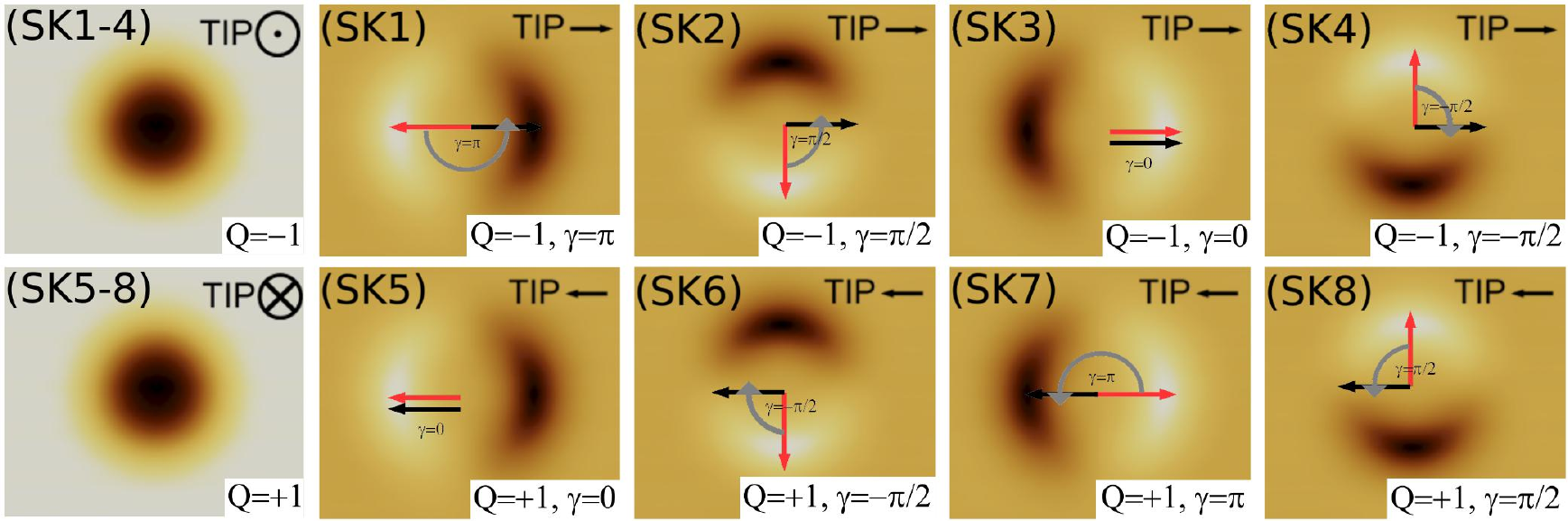}
\end{tabular}
\caption{\label{Fig3} Constant-current SP-STM images of the skyrmions in Fig.\ \ref{Fig1}:
(SK1-4) using an out-of-plane magnetized tip pointing to the $+z$ $[111]$ direction, where skyrmions SK1-4 in Fig.\ \ref{Fig1} show the same contrast; using an in-plane magnetized tip pointing to the $+x$ $[1\bar{1}0]$ direction, where different contrasts are obtained for the skyrmions SK1-4;
(SK5-8) using an out-of-plane magnetized tip pointing to the $-z$ $[\bar{1}\bar{1}\bar{1}]$ direction, where skyrmions SK5-8 in Fig.\ \ref{Fig1} show the same contrast; using an in-plane magnetized tip pointing to the $-x$ $[\bar{1}10]$ direction, where different contrasts are obtained for the skyrmions SK5-8.
Bright and dark contrast respectively means higher and lower apparent height of the constant-current contour. The topological charges ($Q$) and helicities ($\gamma$) of the skyrmions are explicitly shown. The determination of $\gamma$ of the corresponding skyrmion is illustrated in each image (see text for details).}
\end{figure*}

Figure \ref{Fig3} shows SP-STM images of the charge current for the different skyrmions in Fig.\ \ref{Fig1}. The first image on the left hand side of each row in Fig.\ \ref{Fig3} corresponds to out-of-plane magnetized tips, and they show the same contrast for SK1-4 with a $+z$-oriented magnetic tip, and for SK5-8 with a $-z$-oriented magnetic tip. Similarly, the time-reversed spin structure pairs (SK1-SK5, SK2-SK6, SK3-SK7, and SK4-SK8) show the same contrast each with oppositely ($+x$- and $-x$-) oriented in-plane magnetized tips. These findings result from the combined effect of the reversal of the external out-of-plane magnetic field (important for the out-of-plane contrast) and the rotation of the helicity of the skyrmions by $\pi$ (important for the in-plane contrast) during the time-reversal transformation. The in-plane magnetic contrasts in each row from the left to the right are rotating by $\pi/2$ corresponding to the rotation of the helicities of the spin structures by $-\pi/2$ in the series of SK1-4 and SK5-8, separately, thus the in-plane contrast rotation is in anti-phase with the helicity change independently of the sign of $Q$ for the given $m=1$. This finding together with the opposite signs of the topological charge in the series of SK1-4 ($Q=-1$) and the time-reversed SK5-8 ($Q=1$) spin structures suggest that the vorticity $m$ is decisive for the evaluation of the topology of skyrmionic structures from in-plane magnetic contrasts measured by SP-STM, and not the topological charge. Similar findings are expected for arbitrary vorticity values by performing the time-reversal transformation (reversing the external magnetic field), which always changes the sign of the topological charge but leaves the vorticity unchanged. Note that qualitatively the same type of in-plane contrast rotation is observed for the SK1 structure upon rotating the in-plane magnetization direction of the tip \cite{palotas17prb}. In such a case the contrast rotation is expected to be in phase with the tip magnetization rotation for all skyrmions with $m=1$ in Fig.\ \ref{Fig1}. This behavior makes the identification of the helicity of a skyrmion impossible by the evaluation of a measured series of SP-STM images with rotated in-plane-magnetized tips. Instead, the helicity can be determined from a single SP-STM image with a well-defined in-plane tip magnetization orientation: the helicity is the signed angle (represented by gray circular arrow) between the axis pointing from the minimum to the maximum of the two-lobes contrast (red arrow) and the magnetization direction of the tip (black arrow). For illustrating this see Fig.\ \ref{Fig3} for each considered skyrmion. This helicity determination is valid taking a positive value of the effective spin polarization $P_{\mathrm{eff}}$. For a negative $P_{\mathrm{eff}}$ value, the red arrow has to point to the opposite direction, i.e., from the maximum to the minimum of the two-lobes contrast, and the helicity can be obtained similarly as described above.

\subsection{Spin transport characteristics of the skyrmions}
\label{sec_spintr}

\begin{figure*}[t]
\begin{tabular}{c}
\includegraphics[width=1.0\textwidth,angle=0]{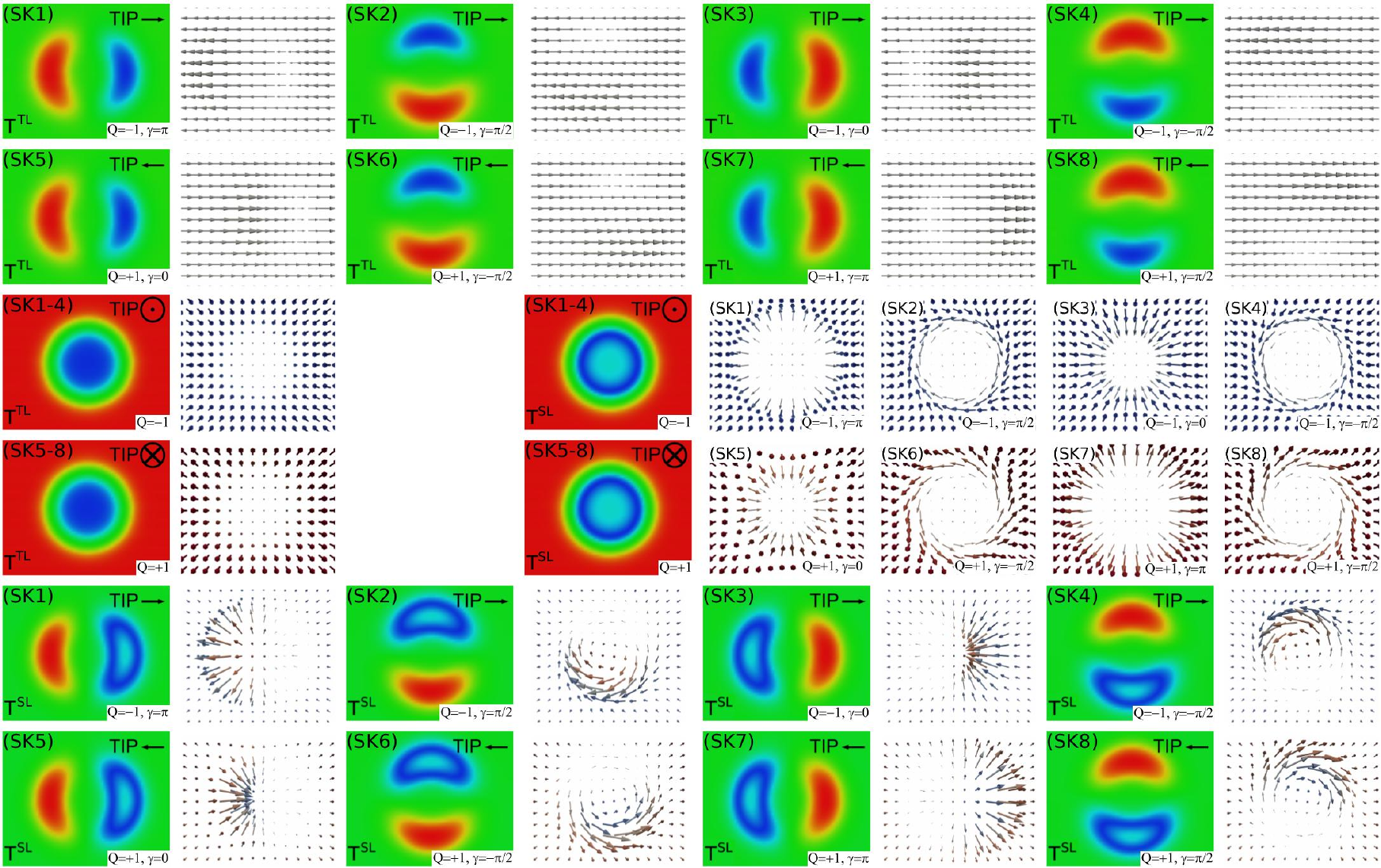}
\end{tabular}
\caption{\label{Fig4} Longitudinal spin current (LSC) magnitudes (red: maximum, blue: minimum) and vectors acting on the scanning tip ($|\mathbf{T}^{TL}|$ and $\mathbf{T}^{TL}$) and on the skyrmions shown in Fig.\ \ref{Fig1} ($|\mathbf{T}^{SL}|$ and $\mathbf{T}^{SL}$) using in-plane and out-of-plane magnetized tips (magnetization directions explicitly indicated) at 6 \AA\;tip-sample distance. Red and blue colors of the LSC vectors respectively correspond to positive and negative out-of-plane ($z$) vector components. The absolute maximal LSC magnitudes are 5.7 neV. The topological charges ($Q$) and helicities ($\gamma$) of the corresponding skyrmions are explicitly shown.}
\end{figure*}

Figure \ref{Fig4} shows calculated longitudinal spin current (LSC) magnitudes and vectors for the different skyrmions in Fig.\ \ref{Fig1}. The top and bottom parts of Fig.\ \ref{Fig4} respectively contain LSC data acting on the tip ($\mathbf{T}^{TL}$) and on the sample ($\mathbf{T}^{SL}$) employing in-plane magnetized tips. The middle part shows calculated LSC quantities with out-of-plane magnetized tips for both $\mathbf{T}^{TL}$ and $\mathbf{T}^{SL}$. In comparison with Fig.\ \ref{Fig3} it is found that the LSC magnitudes show qualitatively the same contrast as the corresponding charge current, and the reasons behind were analyzed in detail in Ref.\ \onlinecite{palotas18stt1}. This implies that the LSC magnitudes show the same contrast for SK1-4 with a $+z$-oriented magnetic tip, and for SK5-8 with a $-z$-oriented magnetic tip. Considering respectively $+x$- and $-x$-oriented in-plane magnetized tips, the time-reversed spin structure pairs (SK1-SK5, SK2-SK6, SK3-SK7, and SK4-SK8) show pairwise the same contrasts for the LSC magnitude as well. Furthermore, an anti-phase in-plane LSC contrast rotation with the helicity change of the skyrmions is clearly visible in Fig.\ \ref{Fig4}. Note that the in-plane LSC contrast rotations are expected to be in phase with the tip magnetization rotations for all skyrmions with $m=1$ in Fig.\ \ref{Fig1}.

The LSC vector maps $\mathbf{T}^{jL}$ with $j\in\{T,S\}$ in Fig.\ \ref{Fig4} on the right hand side of each LSC magnitude maps refer to $T\rightarrow S$ tunneling direction, i.e., to positive bias voltage. According to Eqs.~(\ref{Eq_transport2}), $-\mathbf{T}^{jL}$ refer to LSC vectors of the \emph{same} spin structure at $S\rightarrow T$ tunneling direction, i.e., at negative bias voltage. Apart from the same observed LSC magnitudes for the time-reversed spin structure pairs (SK1-SK5, SK2-SK6, SK3-SK7, and SK4-SK8), by a close inspection of Fig.\ \ref{Fig4}, the following relations can be identified for the LSC vectors:
\begin{eqnarray}
-\mathbf{T}^{jL}(SK1,\gamma=\pi,\mathbf{s}_T)&=&\mathbf{T}^{jL}(SK5,\gamma=0,-\mathbf{s}_T),\nonumber\\
-\mathbf{T}^{jL}(SK2,\gamma=\pi/2,\mathbf{s}_T)&=&\mathbf{T}^{jL}(SK6,\gamma=-\pi/2,-\mathbf{s}_T),\nonumber\\
-\mathbf{T}^{jL}(SK3,\gamma=0,\mathbf{s}_T)&=&\mathbf{T}^{jL}(SK7,\gamma=\pi,-\mathbf{s}_T),\nonumber\\
-\mathbf{T}^{jL}(SK4,\gamma=-\pi/2,\mathbf{s}_T)&=&\mathbf{T}^{jL}(SK8,\gamma=\pi/2,-\mathbf{s}_T),\nonumber\\
\label{Eq_LSC_pattern1}
\end{eqnarray}
where the skyrmion labels in Fig.\ \ref{Fig1} (SK1-8) and their helicities ($\gamma$) are indicated, and the tip magnetization direction $\mathbf{s}_T$ is either out-of-plane ($+z$) or in-plane ($+x$) in the considered cases. Eqs.~(\ref{Eq_LSC_pattern1}) imply that the $\mathbf{T}^{jL}$ vector maps of the SK5-8 set of skyrmions (in every even row of Fig.\ \ref{Fig4}) correspond to the LSC vector maps of the SK1-4 set of skyrmions at $S\rightarrow T$ tunneling direction, i.e., at negative bias voltage, for example $\mathbf{T}^{S\rightarrow T,jL}(SK1,\mathbf{s}_T)=\mathbf{T}^{jL}(SK5,-\mathbf{s}_T)$, etc. Similarly, the $\mathbf{T}^{jL}$ vector maps of the SK1-4 set of skyrmions (in every odd row of Fig.\ \ref{Fig4}) correspond to the LSC vector maps of the SK5-8 set of skyrmions at $S\rightarrow T$ tunneling direction, for example $\mathbf{T}^{S\rightarrow T,jL}(SK5,-\mathbf{s}_T)=\mathbf{T}^{jL}(SK1,\mathbf{s}_T)$, etc.

\begin{figure*}[t]
\begin{tabular}{c}
\includegraphics[width=1.0\textwidth,angle=0]{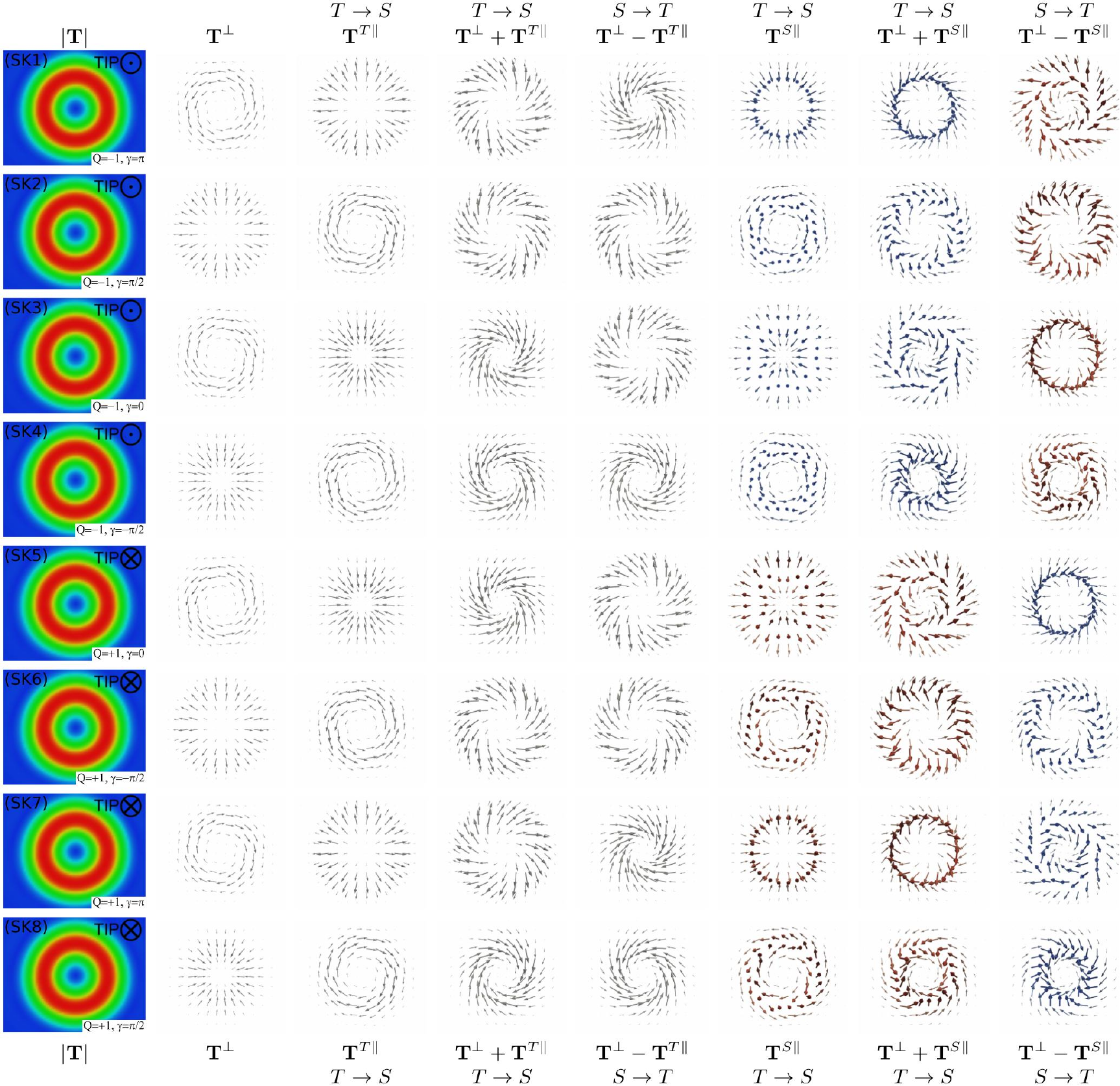}
\end{tabular}
\caption{\label{Fig5} Spin transfer torque (STT) magnitudes $|\mathbf{T}|$ (red: maximum, blue: minimum) and vectors (out-of-plane component ($\mathbf{T}^{\perp}$), in-plane component ($\mathbf{T}^{j\parallel}$), total ($\mathbf{T}^{\perp}\pm\mathbf{T}^{j\parallel}$), depending on the tunneling direction $T\rightarrow S$ or $S\rightarrow T$) acting on the spin moments of the scanning tip ($j=T$) and of the skyrmions ($j=S$) shown in Fig.\ \ref{Fig1} using out-of-plane magnetized tips at 6 \AA\;tip-sample distance. Red and blue colors of the STT vectors respectively correspond to positive and negative out-of-plane ($z$) vector components. The absolute maximal STT magnitudes are 4 neV. The topological charges ($Q$) and helicities ($\gamma$) of the corresponding skyrmions are explicitly shown.}
\end{figure*}

Figure \ref{Fig5} shows calculated spin transfer torque (STT) magnitudes, out-of-plane and in-plane STT vector components, and total STT vectors for the different skyrmions in Fig.\ \ref{Fig1}, employing out-of-plane magnetized tips. Due to the $|\sin\phi_a|$-dependence of the magnitudes of all STT components stemming from the vector products $\mathbf{s}_S^a\times\mathbf{s}_T$ in Eqs.~(\ref{Eq_transport}), the same type of contrast is observed for all STT components independently of the tip magnetization direction \cite{palotas18stt1}, and such a common STT magnitude is denoted by $|\mathbf{T}|$ in Fig.\ \ref{Fig5}. It is clear from the first sight that $|\mathbf{T}|$ show identical contrasts for all skyrmions SK1-8, see the first column of Fig.\ \ref{Fig5}. The STT minima and maxima with out-of-plane magnetized tips are respectively obtained where the tip is placed above regions with dominating out-of-plane and in-plane spin components. This also means that the STT minima are obtained at the regions, where the charge current has minima or maxima \cite{palotas18stt1}. The STT minima show up as blue regions and the maxima as red rings in each image in the first column of Fig.\ \ref{Fig5} since all considered skyrmions in Fig.\ \ref{Fig1} have an axially symmetric shape due to the vorticity of $m=1$. The different helicities do not affect the STT magnitudes employing out-of-plane magnetized tips since the helicities correspond to rotated in-plane spin components only. However, the STT vector components and vectors are sensitive to the helicities of the spin structures. When comparing the STT vector characteristics of the sets of SK1-4 and SK5-8 skyrmions with oppositely magnetized tips, the effects of the time-reversal transformation on both the spin structures ($\mathbf{s}_S^a\rightarrow-\mathbf{s}_S^a$) and the tip magnetization ($\mathbf{s}_T\rightarrow-\mathbf{s}_T$) have to be considered on the STT vector components in Eqs.~(\ref{Eq_transport}).

The calculated STT vectors in Fig.\ \ref{Fig5} show a wide variety depending on the spin moment they are acting on ($T$ or $S$), the skyrmionic structure (SK1-8), and the tunneling direction ($T\rightarrow S$ or $S\rightarrow T$). The out-of-plane STT vectors ($\mathbf{T}^{\perp}$) are components of the total STT vectors acting on both the skyrmionic spins and on the spin of the tip apex, and $\mathbf{T}^{\perp}$ of the \emph{same} spin structure do not depend on the tunneling direction \cite{palotas18stt1}. The following relationships are found concerning $\mathbf{T}^{\perp}$, which lie in the surface plane:
\begin{eqnarray}
\mathbf{T}^{\perp}(SK1,\gamma=\pi,+z)&=&\mathbf{T}^{\perp}(SK5,\gamma=0,-z),\nonumber\\
\mathbf{T}^{\perp}(SK2,\gamma=\pi/2,+z)&=&\mathbf{T}^{\perp}(SK6,\gamma=-\pi/2,-z),\nonumber\\
\mathbf{T}^{\perp}(SK3,\gamma=0,+z)&=&\mathbf{T}^{\perp}(SK7,\gamma=\pi,-z),\nonumber\\
\mathbf{T}^{\perp}(SK4,\gamma=-\pi/2,+z)&=&\mathbf{T}^{\perp}(SK8,\gamma=\pi/2,-z),\nonumber\\
\label{Eq_STT_pattern1perp}
\end{eqnarray}
and
\begin{eqnarray}
\mathbf{T}^{\perp}(SK1,\gamma=\pi,+z)&=&-\mathbf{T}^{\perp}(SK3,\gamma=0,+z),\nonumber\\
\mathbf{T}^{\perp}(SK2,\gamma=\pi/2,+z)&=&-\mathbf{T}^{\perp}(SK4,\gamma=-\pi/2,+z),\nonumber\\
\mathbf{T}^{\perp}(SK5,\gamma=0,-z)&=&-\mathbf{T}^{\perp}(SK7,\gamma=\pi,-z),\nonumber\\
\mathbf{T}^{\perp}(SK6,\gamma=-\pi/2,-z)&=&-\mathbf{T}^{\perp}(SK8,\gamma=\pi/2,-z).\nonumber\\
\label{Eq_STT_pattern2perp}
\end{eqnarray}
Eqs.~(\ref{Eq_STT_pattern1perp}) mean that the time-reversed spin structure pairs (SK1-SK5, SK2-SK6, SK3-SK7, and SK4-SK8) respectively show the same $\mathbf{T}^{\perp}$ with reversed out-of-plane tip magnetization orientations. Similarly, Eqs.~(\ref{Eq_STT_pattern2perp}) mean that the helicity rotation of the spin structures by $\pi$ result in the opposite signs of $\mathbf{T}^{\perp}$. Combining Eqs.~(\ref{Eq_STT_pattern1perp}) and (\ref{Eq_STT_pattern2perp}), the connections between the $\mathbf{T}^{\perp}$ vectors for the skyrmions with the same helicities can be established:
\begin{eqnarray}
\mathbf{T}^{\perp}(SK1,\gamma=\pi,+z)&=&-\mathbf{T}^{\perp}(SK7,\gamma=\pi,-z),\nonumber\\
\mathbf{T}^{\perp}(SK2,\gamma=\pi/2,+z)&=&-\mathbf{T}^{\perp}(SK8,\gamma=\pi/2,-z),\nonumber\\
\mathbf{T}^{\perp}(SK3,\gamma=0,+z)&=&-\mathbf{T}^{\perp}(SK5,\gamma=0,-z),\nonumber\\
\mathbf{T}^{\perp}(SK4,\gamma=-\pi/2,+z)&=&-\mathbf{T}^{\perp}(SK6,\gamma=-\pi/2,-z).\nonumber\\
\label{Eq_STT_pattern3perp}
\end{eqnarray}

For the STT vectors acting on the tip, both components ($\mathbf{T}^{\perp}$ and $\mathbf{T}^{T\parallel}$) and, thus, the total STT vectors ($\mathbf{T}^{\perp}\pm\mathbf{T}^{T\parallel}$) lie in the surface plane since the tip is magnetized in an out-of-plane direction. We find that the $\mathbf{T}^{\perp}$ and the $\mathbf{T}^{T\parallel}$ vectors rotate in phase with the helicity rotation for the SK1-4 and SK5-8 series of skyrmions, see the second and third columns of Fig.\ \ref{Fig5}. Since the helicity rotation follows $-\pi/2$ in the mentioned spin structure series, the STT vector components show an interesting pattern, their directions are the same for each of the following pairs:
\begin{eqnarray}
\mathbf{T}^{T\parallel}(SK1,\gamma=\pi,+z)&\propto&\mathbf{T}^{\perp}(SK2,\gamma=\pi/2,+z),\nonumber\\
\mathbf{T}^{T\parallel}(SK2,\gamma=\pi/2,+z)&\propto&\mathbf{T}^{\perp}(SK3,\gamma=0,+z),\nonumber\\
\mathbf{T}^{T\parallel}(SK3,\gamma=0,+z)&\propto&\mathbf{T}^{\perp}(SK4,\gamma=-\pi/2,+z),\nonumber\\
\mathbf{T}^{T\parallel}(SK4,\gamma=-\pi/2,+z)&\propto&\mathbf{T}^{\perp}(SK1,\gamma=\pi,+z),\nonumber\\
\mathbf{T}^{T\parallel}(SK5,\gamma=0,-z)&\propto&\mathbf{T}^{\perp}(SK8,\gamma=\pi/2,-z),\nonumber\\
\mathbf{T}^{T\parallel}(SK6,\gamma=-\pi/2,-z)&\propto&\mathbf{T}^{\perp}(SK5,\gamma=0,-z),\nonumber\\
\mathbf{T}^{T\parallel}(SK7,\gamma=\pi,-z)&\propto&\mathbf{T}^{\perp}(SK6,\gamma=-\pi/2,-z),\nonumber\\
\mathbf{T}^{T\parallel}(SK8,\gamma=\pi/2,-z)&\propto&\mathbf{T}^{\perp}(SK7,\gamma=\pi,-z).
\label{Eq_STT_pattern1_z}
\end{eqnarray}
Following Eqs.~(\ref{Eq_STT_pattern1perp})-(\ref{Eq_STT_pattern1_z}), the in-plane STT vectors $\mathbf{T}^{T\parallel}$ obey the relationships:
\begin{eqnarray}
\mathbf{T}^{T\parallel}(SK1,\gamma=\pi,+z)&=&-\mathbf{T}^{T\parallel}(SK5,\gamma=0,-z),\nonumber\\
\mathbf{T}^{T\parallel}(SK2,\gamma=\pi/2,+z)&=&-\mathbf{T}^{T\parallel}(SK6,\gamma=-\pi/2,-z),\nonumber\\
\mathbf{T}^{T\parallel}(SK3,\gamma=0,+z)&=&-\mathbf{T}^{T\parallel}(SK7,\gamma=\pi,-z),\nonumber\\
\mathbf{T}^{T\parallel}(SK4,\gamma=-\pi/2,+z)&=&-\mathbf{T}^{T\parallel}(SK8,\gamma=\pi/2,-z),\nonumber\\
\label{Eq_STT_pattern1Tpar}
\end{eqnarray}
and
\begin{eqnarray}
\mathbf{T}^{T\parallel}(SK1,\gamma=\pi,+z)&=&-\mathbf{T}^{T\parallel}(SK3,\gamma=0,+z),\nonumber\\
\mathbf{T}^{T\parallel}(SK2,\gamma=\pi/2,+z)&=&-\mathbf{T}^{T\parallel}(SK4,\gamma=-\pi/2,+z),\nonumber\\
\mathbf{T}^{T\parallel}(SK5,\gamma=0,-z)&=&-\mathbf{T}^{T\parallel}(SK7,\gamma=\pi,-z),\nonumber\\
\mathbf{T}^{T\parallel}(SK6,\gamma=-\pi/2,-z)&=&-\mathbf{T}^{T\parallel}(SK8,\gamma=\pi/2,-z).\nonumber\\
\label{Eq_STT_pattern2Tpar}
\end{eqnarray}
Eqs.~(\ref{Eq_STT_pattern1Tpar}) mean that the time-reversed spin structure pairs (SK1-SK5, SK2-SK6, SK3-SK7, and SK4-SK8) respectively show the opposite $\mathbf{T}^{T\parallel}$ with reversed out-of-plane tip magnetization orientations, similarly as found for the LSC vectors in Eqs.~(\ref{Eq_LSC_pattern1}). Eqs.~(\ref{Eq_STT_pattern2Tpar}) mean that the helicity rotation of the spin structures by $\pi$ result in the opposite signs of $\mathbf{T}^{T\parallel}$, similarly as found for the out-of-plane STT vectors in Eqs.~(\ref{Eq_STT_pattern2perp}). These result in the same $\mathbf{T}^{T\parallel}$ vectors for the spin structures with the same helicities:
\begin{eqnarray}
\mathbf{T}^{T\parallel}(SK1,\gamma=\pi,+z)&=&\mathbf{T}^{T\parallel}(SK7,\gamma=\pi,-z),\nonumber\\
\mathbf{T}^{T\parallel}(SK2,\gamma=\pi/2,+z)&=&\mathbf{T}^{T\parallel}(SK8,\gamma=\pi/2,-z),\nonumber\\
\mathbf{T}^{T\parallel}(SK3,\gamma=0,+z)&=&\mathbf{T}^{T\parallel}(SK5,\gamma=0,-z),\nonumber\\
\mathbf{T}^{T\parallel}(SK4,\gamma=-\pi/2,+z)&=&\mathbf{T}^{T\parallel}(SK6,\gamma=-\pi/2,-z).\nonumber\\
\label{Eq_STT_pattern3Tpar}
\end{eqnarray}

Combining Eqs.\ (\ref{Eq_STT_pattern1perp})-(\ref{Eq_STT_pattern3Tpar}) result in qualitatively similar composing STT vector terms in the $T\rightarrow S$ tunneling direction (at positive bias voltage) for the SK1-SK6, SK2-SK7, SK3-SK8, and SK4-SK5 pairs, each with ($\gamma$)-($\gamma+\pi/2$) helicities, where the $\mathbf{T}^{T\parallel}$ vectors of one structure and the $\mathbf{T}^{\perp}$ vectors of the other structure mutually point to the same direction for each pair, e.g.,\\
$\mathbf{T}^{T\parallel}(SK1,\gamma=\pi,+z)\propto\mathbf{T}^{\perp}(SK6,\gamma=-\pi/2,-z)$ and\\
$\mathbf{T}^{T\parallel}(SK6,\gamma=-\pi/2,-z)\propto\mathbf{T}^{\perp}(SK1,\gamma=\pi,+z)$, etc. Similarly, for the $S\rightarrow T$ tunneling direction (at negative bias voltage) the SK1-SK8, SK2-SK5, SK3-SK6, and SK4-SK7 pairs, each with ($\gamma$)-($\gamma-\pi/2$) helicities, exhibit $-\mathbf{T}^{T\parallel}$ vectors of one structure and $\mathbf{T}^{\perp}$ vectors of the other structure mutually pointing to the same direction for each pair, e.g.,\\
$-\mathbf{T}^{T\parallel}(SK1,\gamma=\pi,+z)\propto\mathbf{T}^{\perp}(SK8,\gamma=\pi/2,-z)$ and\\
$-\mathbf{T}^{T\parallel}(SK8,\gamma=\pi/2,-z)\propto\mathbf{T}^{\perp}(SK1,\gamma=\pi,+z)$, etc.
Thus, the total STT vectors in the mentioned pairs and tunneling directions are qualitatively similar when summing up the components $\mathbf{T}^{\perp}$ and $\pm\mathbf{T}^{T\parallel}$, see the fourth and fifth columns of Fig.\ \ref{Fig5}. However, due to the different prefactors $P_S=-0.5$ and $P_SP_T=0.4$ respectively for $\mathbf{T}^{T\parallel}$ and $\mathbf{T}^{\perp}$ (see Eqs.~(\ref{Eq_transport})), the total STT vectors slightly differ in quantitative terms in the indicated pairs. Exact agreement of the total STT vectors in the SK1-SK6, SK2-SK7, SK3-SK8, and SK4-SK5 pairs at $T\rightarrow S$ tunneling direction ($\mathbf{T}^{\perp}+\mathbf{T}^{T\parallel}$) and in the SK1-SK8, SK2-SK5, SK3-SK6, and SK4-SK7 pairs at $S\rightarrow T$ tunneling direction ($\mathbf{T}^{\perp}-\mathbf{T}^{T\parallel}$) can only be achieved in the ideal case of maximally spin-polarized tips, where $P_T=\pm1$.

The STT vectors acting on the spins of the skyrmions look even more complex in Fig.\ \ref{Fig5}, for example the $\mathbf{T}^{S\parallel}$ vectors are not in the surface plane. The reason is that the $\mathbf{T}^{S\parallel}$ vectors are always in the local $\mathbf{s}_S^a-\mathbf{s}_T$ planes perpendicular to $\mathbf{s}_S^a$, which show a noncollinear feature in the skyrmionic spin structures. We find that the $\mathbf{T}^{S\parallel}$ (sixth column of Fig.\ \ref{Fig5}) and, thus, the total STT vectors at $T\rightarrow S$ tunneling direction ($\mathbf{T}^{\perp}+\mathbf{T}^{S\parallel}$, seventh column of Fig.\ \ref{Fig5}) have negative out-of-plane components ($-z$) for the SK1-SK4 structures with a $+z$-oriented magnetic tip and positive out-of-plane components ($+z$) for the SK5-SK8 structures with a $-z$-oriented magnetic tip. Similarly, the $-\mathbf{T}^{S\parallel}$ vectors and the total STT vectors at $S\rightarrow T$ tunneling direction ($\mathbf{T}^{\perp}-\mathbf{T}^{S\parallel}$, eighth column of Fig.\ \ref{Fig5}) show reversed signs of the out-of-plane vector components compared to the sixth or seventh column of Fig.\ \ref{Fig5}, respectively. These are related to the negative sign of $P_T=-0.8$ taken in the $\mathbf{T}^{S\parallel}$ formula in Eqs.~(\ref{Eq_transport}). It was established \cite{palotas18stt1} that the total $\mathbf{T}^S$ vectors having a positive $z$-component tend to annihilate the skyrmion since in that case the torques would rotate the spins outwards from the surface. This scenario is obtained at $S\rightarrow T$ tunneling direction for SK1-4 skyrmions with $Q=-1$, and at $T\rightarrow S$ tunneling for SK5-8 skyrmions with $Q=1$ with the employed spin polarization parameters. Note that these conclusions crucially depend on the sign of $P_T$ \cite{palotas18stt1}. Moreover, the time-reversed spin structure pairs (SK1-SK5, SK2-SK6, SK3-SK7, and SK4-SK8) show the opposite $\mathbf{T}^{S\parallel}$ vectors each with oppositely out-of-plane magnetized tips, similarly to $\mathbf{T}^{T\parallel}$ in Eqs.~(\ref{Eq_STT_pattern1Tpar}) and to the LSC vectors in Eqs.~(\ref{Eq_LSC_pattern1}). The $\mathbf{T}^{S\parallel}$ vectors are mirror-symmetric to the $xy$-plane (which is perpendicular to the tip magnetization orientation) for the skyrmions with the same helicities:
\begin{eqnarray}
T^{S\parallel}_z(SK1,\gamma=\pi,+z)&=&-T^{S\parallel}_z(SK7,\gamma=\pi,-z),\nonumber\\
T^{S\parallel}_z(SK2,\gamma=\pi/2,+z)&=&-T^{S\parallel}_z(SK8,\gamma=\pi/2,-z),\nonumber\\
T^{S\parallel}_z(SK3,\gamma=0,+z)&=&-T^{S\parallel}_z(SK5,\gamma=0,-z),\nonumber\\
T^{S\parallel}_z(SK4,\gamma=-\pi/2,+z)&=&-T^{S\parallel}_z(SK6,\gamma=-\pi/2,-z).\nonumber\\
\label{Eq_STT_pattern4Tpar}
\end{eqnarray}
Note that Eqs.~(\ref{Eq_STT_pattern3Tpar}) are the special case of this symmetry.

\begin{figure*}[t]
\begin{tabular}{c}
\includegraphics[width=1.0\textwidth,angle=0]{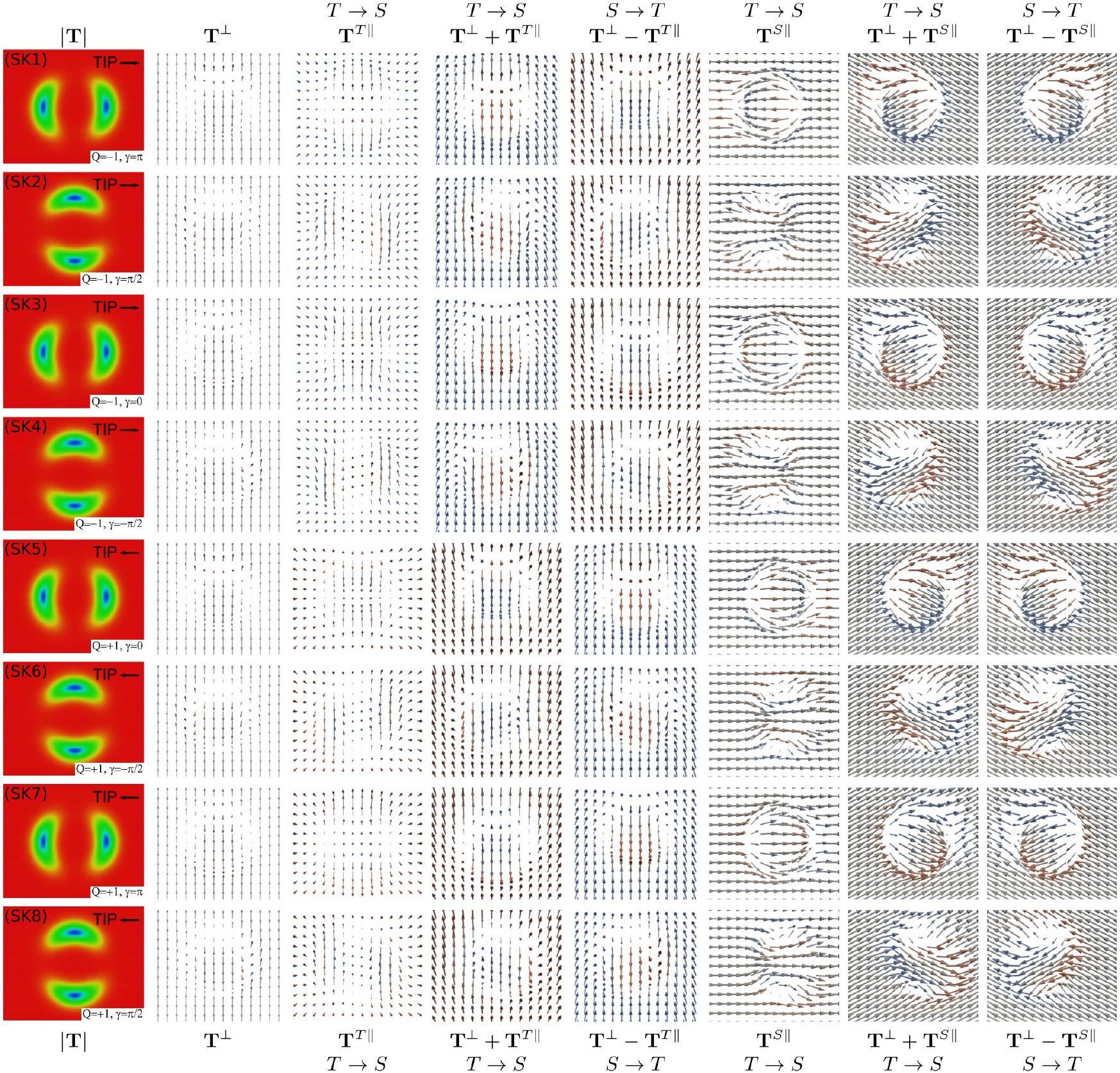}
\end{tabular}
\caption{\label{Fig6} Spin transfer torque (STT) magnitudes $|\mathbf{T}|$ (red: maximum, blue: minimum) and vectors (out-of-plane component ($\mathbf{T}^{\perp}$), in-plane component ($\mathbf{T}^{j\parallel}$), total ($\mathbf{T}^{\perp}\pm\mathbf{T}^{j\parallel}$), depending on the tunneling direction $T\rightarrow S$ or $S\rightarrow T$) acting on the spin moments of the scanning tip ($j=T$) and of the skyrmions ($j=S$) shown in Fig.\ \ref{Fig1} using in-plane magnetized tips at 6 \AA\;tip-sample distance. Red and blue colors of the STT vectors respectively correspond to positive and negative out-of-plane ($z$) vector components. The absolute maximal STT magnitudes are 4 neV. The topological charges ($Q$) and helicities ($\gamma$) of the corresponding skyrmions are explicitly shown.}
\end{figure*}

Figure \ref{Fig6} shows calculated STT magnitudes, out-of-plane and in-plane STT vector components, and total STT vectors for the different skyrmions in Fig.\ \ref{Fig1}, employing in-plane magnetized tips. The magnitudes of all STT components, $|\mathbf{T}|$, show separately the same type of contrast for the N\'eel-skyrmions (SK1, SK3, SK5, SK7: $\gamma=0$ or $\pi$) and for the Bloch-skyrmions (SK2, SK4, SK6, SK8: $\gamma=\pi/2$ or $-\pi/2$), see the first column of Fig.\ \ref{Fig6}. The STT minima (blue) and maxima (red) are respectively obtained where the spins of the skyrmions are in line (parallel or antiparallel) with and perpendicular to the in-plane tip magnetization direction. The observed STT contrasts are closely related to the corresponding in-plane charge current contrasts in Fig.\ \ref{Fig3}. The STT minima correlate very well with the maxima and minima of the charge current \cite{palotas18stt1}, and the reversal of the tip magnetization orientations would not change the STT magnitude images in the first column of Fig.\ \ref{Fig6}. Moreover, the identified anti-phase in-plane charge current contrast rotation with the helicity change in Fig.\ \ref{Fig3} can be transferred to the in-plane contrast rotation of the STT magnitude as well, and the in-plane STT contrast rotations are expected to be in phase with the tip magnetization rotations for all skyrmions with $m=1$ in Fig.\ \ref{Fig1}.

The calculated STT vectors in Fig.\ \ref{Fig6} again show a wide variety depending on the spin moment they are acting on ($T$ or $S$), the skyrmionic structure (SK1-8), and the tunneling direction ($T\rightarrow S$ or $S\rightarrow T$). However, by changing the tip magnetization from out-of-plane to in-plane, the identified relationships of the STT vector components based on Fig.\ \ref{Fig5} are not completely the same since the overall symmetry of the coupled surface-tip system is reduced. As expected, the time-reversed spin structure pairs (SK1-SK5, SK2-SK6, SK3-SK7, and SK4-SK8), each with oppositely in-plane magnetized tips, show the same $\mathbf{T}^{\perp}$ vectors (second column of Fig.\ \ref{Fig6}), similarly to Eqs.~(\ref{Eq_STT_pattern1perp}), and the opposite $\mathbf{T}^{T\parallel}$ and $\mathbf{T}^{S\parallel}$ vectors (third and sixth column of Fig.\ \ref{Fig6}, respectively), similarly to Eqs.~(\ref{Eq_STT_pattern1Tpar}) and to the LSC vectors in Eqs.~(\ref{Eq_LSC_pattern1}). Moreover, the $\mathbf{T}^{\perp}$ and $\mathbf{T}^{T\parallel}$ vectors are respectively mirror-symmetric to the $xy$-plane for the skyrmions with the same helicities:
\begin{eqnarray}
T^{\perp}_z(SK1,\gamma=\pi,+x)&=&-T^{\perp}_z(SK7,\gamma=\pi,-x),\nonumber\\
T^{\perp}_z(SK2,\gamma=\pi/2,+x)&=&-T^{\perp}_z(SK8,\gamma=\pi/2,-x),\nonumber\\
T^{\perp}_z(SK3,\gamma=0,+x)&=&-T^{\perp}_z(SK5,\gamma=0,-x),\nonumber\\
T^{\perp}_z(SK4,\gamma=-\pi/2,+x)&=&-T^{\perp}_z(SK6,\gamma=-\pi/2,-x).\nonumber\\
T^{T\parallel}_z(SK1,\gamma=\pi,+x)&=&-T^{T\parallel}_z(SK7,\gamma=\pi,-x),\nonumber\\
T^{T\parallel}_z(SK2,\gamma=\pi/2,+x)&=&-T^{T\parallel}_z(SK8,\gamma=\pi/2,-x),\nonumber\\
T^{T\parallel}_z(SK3,\gamma=0,+x)&=&-T^{T\parallel}_z(SK5,\gamma=0,-x),\nonumber\\
T^{T\parallel}_z(SK4,\gamma=-\pi/2,+x)&=&-T^{T\parallel}_z(SK6,\gamma=-\pi/2,-x).\nonumber\\
\label{Eq_STT_pattern1_x}
\end{eqnarray}
These mean that the total STT vectors acting on the tip, $\mathbf{T}^{\perp}\pm\mathbf{T}^{T\parallel}$ (fourth and fifth column of Fig.\ \ref{Fig6}, respectively), are also mirror-symmetric to the $xy$-plane for the skyrmions with the same helicities. Furthermore, it is found that the $\mathbf{T}^{S\parallel}$ vectors are mirror-symmetric to the $yz$-plane (which is perpendicular to the tip magnetization orientation) for the spin structures with the same helicities of the Bloch-skyrmions and with $\pi$-rotated helicities of the N\'eel-skyrmions:
\begin{eqnarray}
T^{S\parallel}_x(SK1,\gamma=\pi,+x)&=&-T^{S\parallel}_x(SK5,\gamma=0,-x),\nonumber\\
T^{S\parallel}_x(SK2,\gamma=\pi/2,+x)&=&-T^{S\parallel}_x(SK8,\gamma=\pi/2,-x),\nonumber\\
T^{S\parallel}_x(SK3,\gamma=0,+x)&=&-T^{S\parallel}_x(SK7,\gamma=\pi,-x),\nonumber\\
T^{S\parallel}_x(SK4,\gamma=-\pi/2,+x)&=&-T^{S\parallel}_x(SK6,\gamma=-\pi/2,-x).\nonumber\\
\label{Eq_STT_pattern2_x}
\end{eqnarray}
The same symmetry can be observed for the total STT vectors acting on the skyrmionic spins, $\mathbf{T}^{\perp}\pm\mathbf{T}^{S\parallel}$ (seventh and eighth column of Fig.\ \ref{Fig6}, respectively). It is expected that correspondingly similar mirror-symmetries for the STT vector components apply with respect to the perpendicular plane to an arbitrary in-plane tip magnetization orientation.

It is found that the following relationships for the STT vectors hold independently of the tip magnetization orientation:
\begin{eqnarray}
\mathbf{T}^{T\rightarrow S,j}(SK1,\mathbf{s}_T)&=&\mathbf{T}^{S\rightarrow T,j}(SK5,-\mathbf{s}_T),\nonumber\\
\mathbf{T}^{T\rightarrow S,j}(SK2,\mathbf{s}_T)&=&\mathbf{T}^{S\rightarrow T,j}(SK6,-\mathbf{s}_T),\nonumber\\
\mathbf{T}^{T\rightarrow S,j}(SK3,\mathbf{s}_T)&=&\mathbf{T}^{S\rightarrow T,j}(SK7,-\mathbf{s}_T),\nonumber\\
\mathbf{T}^{T\rightarrow S,j}(SK4,\mathbf{s}_T)&=&\mathbf{T}^{S\rightarrow T,j}(SK8,-\mathbf{s}_T)\nonumber\\
\mathbf{T}^{S\rightarrow T,j}(SK1,\mathbf{s}_T)&=&\mathbf{T}^{T\rightarrow S,j}(SK5,-\mathbf{s}_T),\nonumber\\
\mathbf{T}^{S\rightarrow T,j}(SK2,\mathbf{s}_T)&=&\mathbf{T}^{T\rightarrow S,j}(SK6,-\mathbf{s}_T),\nonumber\\
\mathbf{T}^{S\rightarrow T,j}(SK3,\mathbf{s}_T)&=&\mathbf{T}^{T\rightarrow S,j}(SK7,-\mathbf{s}_T),\nonumber\\
\mathbf{T}^{S\rightarrow T,j}(SK4,\mathbf{s}_T)&=&\mathbf{T}^{T\rightarrow S,j}(SK8,-\mathbf{s}_T),
\label{Eq_STT_pattern5}
\end{eqnarray}
where $j\in\{T,S\}$, and $\mathbf{s}_T$ is either out-of-plane ($+z$) or in-plane ($+x$) in the considered cases, see Fig.\ \ref{Fig5} and \ref{Fig6}, respectively. Eqs.~(\ref{Eq_STT_pattern5}) establish the connections between the total STT vectors of the time-reversed spin structure pairs at opposite tip magnetizations and tunneling directions.

As a final note, the experimental realization of the measurement of vector spin transport quantities could be along the following line: Sankey et al.\ \cite{sankey08} reported the first direct measurement of the out-of-plane and in-plane components of the STT vector in a planar magnetic tunnel junction without spatial resolution. In these experiments the relevant information was obtained from ferromagnetic resonance (FMR) peaks. Recently, a spatially resolved FMR experimental tool attached to an SP-STM was reported \cite{herve17conf}. This technique is very promising for the future measurement of STT vector components with spatial resolution in SP-STM.

\section{Summary and conclusion}
\label{sec_conc}

Based on a combined tunneling electron charge and vector spin transport theory capable of imaging noncollinear magnetic textures on surfaces with spin-polarized scanning tunneling microscopy (SP-STM), the high-resolution charge and spin transport properties of a variety of N\'eel- and Bloch-type skyrmions were investigated. Axially symmetric skyrmions were considered within the same topology class characterized by a vorticity value of $m=1$, and their helicities were varied taking $\gamma=0$ and $\pi$ values for the N\'eel skyrmions and $\gamma=\pm\pi/2$ values for the Bloch skyrmions. The effect of the time-reversal transformation on the spin structures was also considered, altogether resulting in eight skyrmions under study.

It was demonstrated that single SP-STM images can be used to uniquely determine the helicity of the skyrmions once the orientation of the tip magnetization and the effective spin polarization are known. It was found that the in-plane SP-STM contrast rotation is in anti-phase with the helicity change of the skyrmions. Furthermore, it was suggested that the vorticity $m$ is decisive for the evaluation of the topology of skyrmionic structures from in-plane charge current contrasts measured by SP-STM, and not the topological charge $Q$ since $m$ does not change sign upon time-reversal transformation (by reversing the external magnetic field), while $Q$ does.

The magnitudes of the current-induced spin transport vector quantities, the longitudinal spin current and the spin transfer torque, show close relation to standard charge current SP-STM images. Depending on the orientation of the tip magnetization as well as on the helicity and the time-reversal of the skyrmionic spin structures, several relationships between their spin transport vector components, the in-plane and out-of-plane spin transfer torque (STT) and the longitudinal spin current, were identified. For example, both STT vector components rotate in phase with the helicity rotation for the SK1-4 and the time-reversed SK5-8 series of skyrmions using out-of-plane magnetized tips.

As a further development of the 3D-WKB electron tunneling theory, the modified spin polarization vectors of the conduction electrons due to the local non-coplanarity (chirality) of the complex spin texture were incorporated into the model. It was found that this effect modifies the apparent size of the skyrmions. This should be considered when evaluating actual skyrmion sizes from SP-STM measurements and related material-specific magnetic interaction and anisotropy parameters.

The obtained results lay the basis for the proper identification of topological surface magnetic objects with different helicities imaged by SP-STM, and deliver important vector spin transport parameters for current-induced spin dynamics due to local spin-polarized currents in SP-STM.

\section{Acknowledgments}

The author thanks Levente R\'ozsa for the SK1 skyrmion data and discussions. Financial supports of the SASPRO Fellowship of the Slovak Academy of Sciences (project No.\ 1239/02/01), the National Research, Development, and Innovation Office of Hungary under Projects No.\ K115575 and No.\ FK124100, and the BME-Nanotechnology FIKP grant of EMMI (BME FIKP-NAT) are gratefully acknowledged.

\end{document}